\documentclass[journal]{IEEEtran}
\usepackage{cite}
\usepackage{amsmath,amssymb,amsfonts}
\usepackage{graphicx}
\usepackage{textcomp}
\usepackage{csquotes}
\usepackage{xcolor,soul}
\usepackage{adjustbox}
\usepackage{hyperref}
\usepackage{multirow}
\usepackage{titlesec}
\usepackage{subcaption}
\usepackage{makecell}
\usepackage{orcidlink}
\usepackage{booktabs} 
\usepackage{caption} 
\usepackage[numbers]{natbib}
\usepackage{cancel}

\usepackage{etoolbox}
\usepackage{adjustbox}
\usepackage{amsthm}

\usepackage{siunitx}
\usepackage{algorithm}
\usepackage{algpseudocode}

\definecolor{columbiablue}{rgb}{0.61, 0.87, 1.0}
\titlespacing{\subsubsection}{0pt}{0pt}{1pt}

\sethlcolor{white}

\begin{document}

\title{Leveraging Past DVL Velocity Measurements for Acceleration-Aided AUV Navigation}

\author{Nadav~Cohen$^{*}$ \orcidlink{0000-0002-8249-0239},
Orzion~Levy$^{*}$ \orcidlink{0009-0004-2844-1518},
and~Itzik~Klein \orcidlink{0000-0001-7846-0654}
\thanks{$^{*}$Nadav Cohen and Orzion Levy contributed equally to this work.}
\thanks{N. Cohen, O. Levy and I. Klein are with the Hatter Department of Marine Technologies, Charney School of Marine Sciences, University of Haifa, Israel. Corresponding author: ncohe140@campus.haifa.ac.il (N. Cohen).}
}

\maketitle

\begin{abstract}
Autonomous underwater vehicles (AUVs) commonly rely on the fusion of an inertial navigation system (INS) with a Doppler velocity log (DVL) through a nonlinear filter to maintain accurate navigation in underwater environments. In this paper, we propose a cost-free, software-only extension to the standard INS/DVL error-state extended Kalman filter (EKF). The proposed method derives an acceleration estimate from a polynomial fit of past DVL velocity measurements and incorporates it as an additional measurement update. We provide a closed-form derivation of the acceleration measurement model and its associated noise covariance. The proposed approach requires no additional hardware, operates at the existing DVL sampling rate, and is implemented through a one-time software modification. We evaluate our approach using real sea-trial data collected with the Snapir AUV across three trajectories with distinct maneuvering profiles. Compared to the baseline EKF, the proposed method achieves up to 50\% improvement in position accuracy, 59\% in velocity accuracy, and 64\% in orientation accuracy, with the percent covariance improvement metric confirming a genuine reduction in estimator uncertainty rather than a mere shift in the point estimate. These results demonstrate that DVL-derived acceleration measurements offer a practical and low-cost means of enhancing underwater navigation performance across diverse AUV operating conditions.
\end{abstract}


\IEEEpeerreviewmaketitle

\section{Introduction}
\noindent
\IEEEPARstart{E}{xploring} the underwater environment remains a challenge humankind is still trying to tackle. Despite covering more than 70\% of the Earth's surface, most of the oceans remain unexplored and inaccessible using conventional means, driving a growing need for systems that are both affordable and scalable enough to close this gap~\cite{mcintyre2010life,liu2021ocean}. Autonomous underwater vehicles (AUVs) hold the most promise to do so, offering the ability to operate for extended durations in environments too remote, deep, or hazardous for direct human access. In recent years, rather than mounting AUVs with high-end and expensive sensors, the shift has been towards lower-cost sensors that enable a similar level of reliability while being more affordable and scalable, thereby making large-scale deployment of AUVs a more practical and economically viable proposition~\cite{campagnaro2023survey}.
\\ \noindent
The challenge in navigating the underwater environment lies in the fact that global navigation satellite system (GNSS) signals cannot penetrate water and therefore cannot be used to aid navigation. To overcome this difficulty, an inertial navigation system (INS) is commonly fused with a Doppler velocity log (DVL)~\cite{miller2010autonomous,luo2019new,mu2021practical,wu2019survey}. The inertial readings from the inertial sensors can be integrated, and through dead reckoning provide the full navigation solution, consisting of position, velocity, and orientation. However, due to built-in errors within the sensor, this solution tends to accumulate error over time. To mitigate this error, the DVL sensor provides an accurate velocity measurement~\cite{titterton2004strapdown,groves2015principles}. While INS/DVL fusion is the common approach, it still introduces difficulties, one major one of which is the unobservability of position. The common way to fuse the INS and DVL is through a nonlinear filter, such as the extended Kalman filter (EKF) or the unscented Kalman filter (UKF)~\cite{bar2001estimation,simon2006optimal,farrell2008aided,xu2021accurate}.
\\ \noindent
When using a high-end and expensive inertial sensor, such as a fiber-optic one, that is accurate enough to sense Earth's rotation, the accumulated error is not substantial over an AUV's survey time period. But when working with a low-cost IMU, such as a micro-electro-mechanical system (MEMS) one, which is a few orders of magnitude worse, the error is substantial and needs to be addressed. Recent works in the field of AUV navigation try to address this and maintain reliable navigation using information-aided mechanisms, where, for example, zero velocity updates (ZUPT) are incorporated when appropriate~\cite{yao2017imm,yao2022virtual}, as well as other techniques~\cite{engelsman2023information}.
More strategies using data-driven approaches have been suggested to enhance DVL velocity accuracy~\cite{cohen2022beamsnet,zhang2025underwater,zhang2026ins,yampolsky2025dcnet}, improve filter performance by adaptively accommodating a changing environment through the process noise matrix~\cite{cohen2025adaptive,levy2026adaptive,kang2026hybrid}, and overcome cases of partial DVL measurements for seamless underwater navigation~\cite{topini2023experimental,cohen2024seamless}. In~\cite{klein2022estimating}, the authors introduce a method to overcome complete DVL outages in extreme underwater environments, where past DVL measurements are used through a polynomial fit to retrieve the missing measurement.
\\ \noindent
Building upon this work, in this paper we aim to improve INS/DVL fusion through the EKF for AUV navigation without any additional sensor, using only a software implementation. We aim to use the already-acquired information from the DVL, and past measurements to fit a polynomial that can provide the acceleration of the AUV. In doing so, we derive and provide an additional update in the INS/DVL fusion associated with the acceleration bias. This helps improve filter convergence, performance, and certainty. The DVL acceleration vector is a cost-free measurement that requires only a one-time software modification and is available at the DVL sampling rate. Once applied, it demonstrated improvement across all states. Specifically, the primary contributions of this work are as follows:
\begin{enumerate}
    \item A novel method that enhances INS/DVL fusion performance by incorporating acceleration measurements derived from past DVL velocity data.
    \item Development of an analytical, closed-form acceleration measurement model and its associated measurement noise covariance.
    \item The proposed approach requires no additional hardware costs and is easily implemented via a one-time software modification. It operates at the existing DVL sampling rate, ensuring full compatibility with standard real-time systems.
\end{enumerate}

\noindent
To evaluate our proposed approach, we performed sea trials. We demonstrated up to $50\%$ improvement in position accuracy, up to $59\%$ improvement in velocity state, and up to $64\%$ improvement in orientation estimation compared to the baseline approach. In addition, our approach provided better certainty in the estimation and overall navigation performance across three different test scenarios that mimic most of the AUV's maneuvers.\\ \noindent
The remainder of this paper is organized as follows: Section \ref{sec:system} presents the INS/DVL system modeling, including the reference frames and the baseline error-state EKF formulation. Section \ref{sec:proposed} introduces the proposed acceleration-aided update, deriving the acceleration estimation model, the extended DVL update, and an observability analysis of the added measurement. \ref{sec:results} presents the analysis and results, covering the dataset acquisition, data augmentation procedure, evaluation metrics, and a detailed discussion of the experimental results. Finally, Section \ref{sec:conclude} concludes the paper.
\section{INS/DVL System Modeling}\label{sec:system}
\subsection{Reference Frames}
\noindent 
Here we define the reference frames used throughout the derivation of our approach~\cite{britting2010inertial}:
\begin{itemize}
\item \textbf{Earth-Centered Earth-Fixed (ECEF) frame}: A global Cartesian frame centered at the Earth's core. The x-axis passes through the prime meridian at the equator, the z-axis is aligned with the Earth's spin axis, and the y-axis is set to complete a right-handed system.
\item \textbf{Navigation frame}: A local, tangent-plane frame anchored to the platform's current position. We adopt the north-east-down (NED) convention, in which north and east define the x- and y-axes, respectively, while the down-axis is oriented roughly along the gravity vector. The AUV's navigation solution is reported in this frame.
\item \textbf{Body frame}: Defined by the AUV's physical structure under the right-hand rule. The surge axis (x) runs along the vehicle's longitudinal centerline toward the nose, the sway axis (y) points laterally to starboard, and the heave axis (z) points downward. We assume the onboard inertial sensors in the inertial measurement unit (IMU) are aligned with the body frame.
\item \textbf{DVL frame}: The frame in which the DVL sensor reports velocity, defined by the sensor manufacturer rather than the vehicle geometry. Because mounting the DVL is subject to electrical and mechanical constraints, this frame is generally offset from the body frame rather than coincident with it. However, since the DVL-to-body transformation is fixed and known, we assume without loss of generality that the DVL is installed directly in the body frame.
\end{itemize}
\subsection{INS/DVL Fusion}
\noindent Expressed in NED coordinates, the INS equations of motion are given by \cite{farrell2008aided}:
\begin{equation}\label{eq_ins1}
\boldsymbol{\dot{v}}^{n}=\mathbf{R}^{n}_{b}\boldsymbol{f}^{b}_{ib}+\boldsymbol{g}^{n}-\left(\mathbf{\Omega}^{n}_{en}+2\mathbf{\Omega}^{n}_{ie}\right) \boldsymbol{v}^{n}
\end{equation}
\begin{equation}\label{eq_ins2}
\dot{\mathbf{R}}^{n}_{b}=\mathbf{R}^{n}_{b}\mathbf{\Omega}^{b}_{ib}-\left(\mathbf{\Omega}^{n}_{ie}+\mathbf{\Omega}^n_{en}\right)\mathbf{R}^{n}_{b}
\end{equation}
where, $\boldsymbol{v}^{n}$ denotes the navigation-frame velocity vector, $\boldsymbol{g}^{n}$ the navigation-frame gravity vector, $\mathbf{\Omega}^{n}_{en}$ the skew-symmetric matrix of the navigation-frame angular velocity relative to ECEF, and $\mathbf{\Omega}^{n}_{ie}$ the skew-symmetric matrix of Earth's rotation rate. $\mathbf{R}^{n}_{b}$ maps the body frame to the NED frame, $\boldsymbol{f}^{b}_{ib}$ is the measured specific force, and $\mathbf{\Omega}^{b}_{ib}$ is the skew-symmetric matrix of the gyroscope-measured angular rate. \newline
Equations \eqref{eq_ins1}-\eqref{eq_ins2} serve as the nonlinear system dynamics of the filter, giving rise to a 12-dimensional error-state vector $\boldsymbol{\delta x} \in R^{12}$:
\begin{equation} \label{eq:errorState}
\boldsymbol{\delta x} = \left[ \begin{array}{cccc}
\boldsymbol{\delta v^{n}} & \boldsymbol{\phi^{n}} & \boldsymbol{b_{a}} & \boldsymbol{b_{g}} \end{array} \right]^{T} 
\end{equation}
where, $\boldsymbol{\delta v^{n}}$ is the velocity error states expressed in the navigation frame, $\boldsymbol{\phi^{n}}$ the misalignment error states, and $\boldsymbol{b_{a}}$, $\boldsymbol{b_{g}}$ the residual biases of the accelerometer and gyroscope, respectively. We model the inertial residuals as a random-walk process driven by zero-mean white Gaussian noise, $\boldsymbol{w_{a_{b}}}$ for the accelerometer and $\boldsymbol{w_{g_{b}}}$ for the gyroscope.
\\
\noindent
Linearizing yields the following error-state dynamics~\cite{grewal2007global}:
\begin{equation}
\boldsymbol{\delta \dot {x}} =\mathbf{F} \boldsymbol{\delta x} +\mathbf{G}\boldsymbol{w} 
\end{equation}
where $\mathbf{F}$ denotes the system matrix, and can be found in~\cite{farrell2008aided}. $\mathbf{G}$ is the shaping matrix:
\begin{equation}
\boldsymbol{\mathrm{G}} =~\left[ \begin{array}{cccc}
\mathbf{R}^{n}_{b} & \boldsymbol{0_{3\times3}} & \boldsymbol{0_{3\times3}} & \boldsymbol{0_{3\times3}} \\ 
\boldsymbol{0_{3\times3}} & \mathbf{R}^{n}_{b} & \boldsymbol{0_{3\times3}} & \boldsymbol{0_{3\times3}} \\ 
\boldsymbol{0_{3\times3}} & \boldsymbol{0_{3\times3}} & \mathbf{{I}_{3}} & \boldsymbol{0_{3\times3}} \\ 
\boldsymbol{0_{3\times3}} & \boldsymbol{0_{3\times3}} & \boldsymbol{0_{3\times3}} & \mathbf{{I}_{3}} \end{array}
\right],
\end{equation}
and $\boldsymbol{w}$ is the noise vector:
\begin{equation}
\boldsymbol{w} = \left[ \begin{array}{cccc}
\boldsymbol{w_{a}} & \boldsymbol{w_{g}} & \boldsymbol{w_{a_{b}}} & \boldsymbol{w_{g_{b}}} \end{array} \right]^{T}, 
\end{equation}
with $\boldsymbol{w_{a}}$, $\boldsymbol{w_{g}}$ the accelerometer and gyroscope measurement noise vectors, respectively, and $\boldsymbol{w_{a_{b}}}$, $\boldsymbol{w_{g_{b}}}$ are the corresponding accelerometer and gyroscope biases noise terms, respectively.
\newline
The error-state covariance is propagated as:
\begin{equation}\label{eq_p}
\noindent
\mathbf{\hat{P}}^{-}_{k} = \mathbf{\Phi}_{k-1} \mathbf{\hat{P}}^{+}_{k-1}  \mathbf{\Phi}_{k-1}^{T} +  \mathbf{Q}_{k-1}
\end{equation}

\noindent
where $\mathbf{\hat{P}}^{-}_{k}$ and $\mathbf{\hat{P}}^{+}_{k}$ denote the \textit{a-prior}i and \textit{a-posteriori} covariance estimates, respectively, and $\mathbf{Q}$ is the process noise covariance.\newline  
Because the INS equations are nonlinear, the dynamic matrix $\mathbf{\Phi}$ used in \eqref{eq_p} is obtained via a first-order approximation:
\begin{equation}
\mathbf{\Phi} = \mathbf{I} + \mathbf{F} \cdot dt 
\end{equation}
with $dt$ the navigation time step.\newline
The Kalman gain is computed as:
\begin{equation}
\mathbf{K}_{k} = \mathbf{\hat{P}}^{-}_{k}\mathbf{H}_{k}^{T}(\mathbf{H}_{k}\mathbf{\hat{P}}^{-}_{k}\mathbf{H}_{k}^{T} + \mathbf{R}_{k})^{-1}
\end{equation}
where $\mathbf{H}$ and $\mathbf{R}$ denote the measurement matrix and measurement noise covariance, respectively. \\
The \textit{a-posteriori} covariance is then updated:
\begin{equation}
\mathbf{\hat{P}}^{+}_{k} = [\mathbf{I} - \mathbf{K}_{k}\mathbf{H}_{k}]\mathbf{\hat{P}}^{-}_{k}
\end{equation}
and the closed-loop error state follows as \cite{groves2015principles}:
\begin{equation}
\boldsymbol{\delta \hat{x}}^{+}_{k} =  \mathbf{K}_{k} \boldsymbol{\delta z}_{k}
\end{equation}
where $\boldsymbol{\delta \hat{x}}^{+}_{k}$ is the \textit{a-posteriori} error-state estimate and $\boldsymbol{\delta z}$ is the measurement residual vector.\\
The baseline INS/DVL fusion relies solely on DVL velocity measurements, giving a measurement residual of:
\begin{equation} \label{eq: z_vel}
\delta\boldsymbol{z}_v = \boldsymbol{\hat{v}}^{b} - \boldsymbol{\tilde{v}}_{DVL}^{b}  
\end{equation}
where $\boldsymbol{\tilde{v}}_{DVL}^{b}$ is the DVL velocity measurement expressed in the body frame, defined as:
\begin{equation} \label{eq: V_DVL_n}
\boldsymbol{\tilde{v}}_{DVL}^{b} = \mathbf{\hat{R}}_{n}^{b} \boldsymbol{\tilde{v}}_{DVL}^{n},
\end{equation}
and $\boldsymbol{\tilde{v}}_{DVL}^{n}$ is the DVL velocity measurement in navigation-frame axes.
Substituting \eqref{eq: V_DVL_n} into \eqref{eq: z_vel} yields:
\begin{equation}
\boldsymbol{\delta\tilde{v}}_{DVL}^{b} = \mathbf{\hat{R}}_{n}^{b} - 
\mathbf{\hat{R}}_{n}^{b}
[\boldsymbol{\tilde{v}}_{DVL}^{n} \times]
\boldsymbol{\phi^{n}}.
\end{equation}
The resulting measurement matrix $\mathbf{H}_v$ is:
\begin{equation}\label{eq:DVL_vel_H}
\mathbf{H}_v = \left[ \begin{array}{cccc}
\mathbf{\hat{R}}_{n}^{b} & -\mathbf{\hat{R}}_{n}^{b}
[\boldsymbol{\tilde{v}}_{DVL}^{n} \times] & \boldsymbol{0_{3\times3}} & \boldsymbol{0_{3\times3}} \end{array} \right]. 
\end{equation}
\section{Proposed Acceleration-Aided Update}\label{sec:proposed}
\begin{figure*}[!ht]
\centering
\includegraphics[width = 0.85\linewidth]{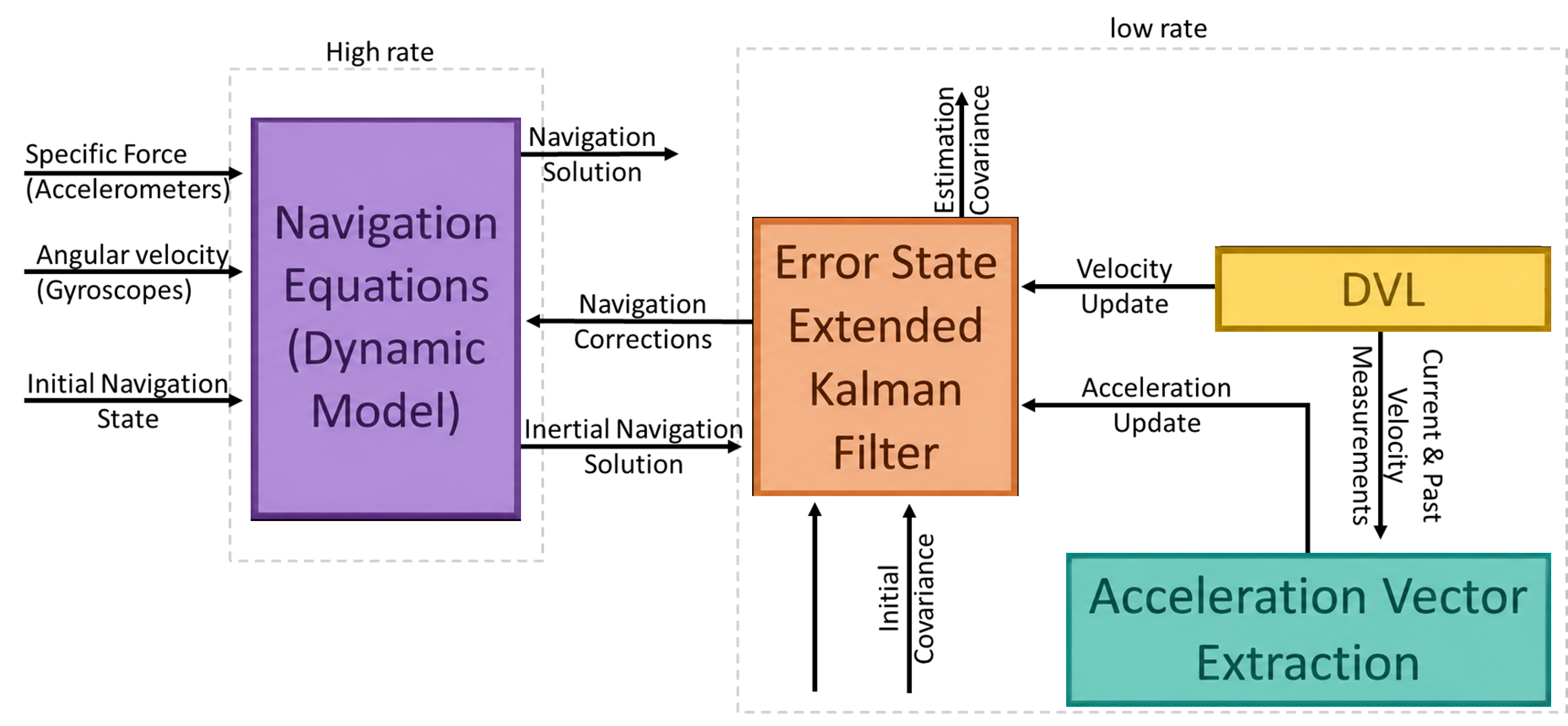} 
\caption{{Flow diagram of our proposed approach embedded within the INS/DVL fusion pipeline. Our approach provides DVL-based acceleration vector updates to the navigation filter. }}
\label{fig: Algorithm Diagram}
\end{figure*}
\noindent
In this section, we introduce the proposed acceleration-aided extension to the baseline INS/DVL fusion described above. Rather than relying solely on the DVL's instantaneous velocity measurement, we leverage short history of past DVL velocity readings to derive an auxiliary acceleration estimate through a polynomial fit. This estimate is then incorporated into the error-state EKF as an additional measurement update, alongside the standard velocity update, providing the filter with richer information about the vehicle's dynamics at no extra hardware cost. We first derive the acceleration estimation model and its associated measurement noise covariance, then formulate the extended DVL update model, and conclude with an observability analysis to assess the added measurement's contribution to the system.
\subsection{Acceleration Estimation} \label{sec:AE}
\noindent
Assuming the velocity is a continuous signal, $\boldsymbol{v}(t) \in \mathbb{R}^3$, it can be locally approximated by a truncated Taylor series about $t_0$~\cite{klein2022estimating}:
\begin{equation}\label{eqn:v_pol}
\boldsymbol{v}(t)
=
\boldsymbol{v}_0
+\dot{\boldsymbol{v}}_0 (t-t_0)
+\frac{1}{2!}\ddot{\boldsymbol{v}}_0 (t-t_0)^2
+\frac{1}{3!}\boldsymbol{v}^{(3)}_0 (t-t_0)^3
+\ldots
\end{equation}
or, retaining $n$ terms, in matrix form:
\begin{equation}\label{eq_poly}
\boldsymbol{v}(t)
=
\underbrace{
\begin{bmatrix}
1 &
(t-t_0) &
\dfrac{(t-t_0)^2}{2!} &
\cdots &
\dfrac{(t-t_0)^{n-1}}{(n-1)!}
\end{bmatrix}}_{\boldsymbol{\psi}^{\mathsf T}(t)}
\underbrace{
\begin{bmatrix}
\boldsymbol{v}_0^{\mathsf T} \\
\dot{\boldsymbol{v}}_0^{\mathsf T} \\
\ddot{\boldsymbol{v}}_0^{\mathsf T} \\
\vdots \\
\boldsymbol{v}^{(n-1)\mathsf T}_0
\end{bmatrix}}_{\boldsymbol{V}} ,
\end{equation}
where $n$ is the number of retained polynomial terms, and $\boldsymbol{V} \in \mathbb{R}^{n\times3}$ stacks the velocity and its time derivatives at $t_0$.
\\ \noindent
Let $m$ past velocity measurements be sampled at times
$t_0,t_1,\ldots,t_{m-1}$. 
\hl{We assume that deterministic DVL errors, including bias, scale-factor, and misalignment errors, have been compensated through sensor calibration. }
Accordingly, each DVL velocity measurement is modeled as the true velocity
corrupted by zero-mean white Gaussian noise:
\begin{equation}\label{eqn:noise_yk}
\boldsymbol{y}_k = \boldsymbol{v}(t_k) + \boldsymbol{\omega}_{v}, \qquad
\boldsymbol{\omega}_v \sim \mathcal{N}(\boldsymbol{0}, \sigma_v^2 \mathbf{I}_3).
\end{equation}
Stacking the measurements and the corresponding basis rows from \eqref{eq_poly} gives:
\begin{equation}
\mathbf{Y} := 
\begin{bmatrix}
\boldsymbol{y}_0^{\mathsf T}\\
\boldsymbol{y}_1^{\mathsf T} \\
\vdots\\
\boldsymbol{y}_{m-1}^{\mathsf T}
\end{bmatrix}_{m\times 3}, 
\qquad
\mathbf{\Psi} := 
\begin{bmatrix}
\boldsymbol{\psi}^{\mathsf T}(t_0) \\
\boldsymbol{\psi}^{\mathsf T}(t_1) \\
\vdots\\
\boldsymbol{\psi}^{\mathsf T}(t_{m-1})
\end{bmatrix}_{m\times n},
\end{equation}
so that the measurement model is $\mathbf{Y} = \mathbf{\Psi}\boldsymbol{V} + \boldsymbol{\Omega}_v$, with $\boldsymbol{\Omega}_v$ the stacked noise matrix.
$\boldsymbol{V}$ is being estimated via ordinary least squares:
\begin{equation}\label{eq_argmin}
\hat{\boldsymbol{V}} = \operatorname*{arg\,min}_{\boldsymbol{V}} \; \|\mathbf{Y}-\mathbf{\Psi}\boldsymbol{V}\|^2 .
\end{equation}
Provided $m \geq n$ and $\mathbf{\Psi}$ has full column rank, the unique solution is:
\begin{equation}\label{eq_Vhat}
\hat{\boldsymbol{V}} = (\mathbf{\Psi}^{\mathsf T}\mathbf{\Psi})^{-1}\mathbf{\Psi}^{\mathsf T}\mathbf{Y}.
\end{equation}
The estimator \eqref{eq_Vhat} is unbiased, and under the white-noise assumption above it is the best linear unbiased estimator (BLUE) by the Gauss--Markov theorem~\cite{bar2001estimation}.
\\ \noindent
 The acceleration at $t_0$ is the second row of $\hat{\boldsymbol{V}}$, i.e., the coefficient of $\dot{\boldsymbol{v}}_0$. Defining the selection vector $\textbf{e}_2 = \begin{bmatrix} 0 & 1 & 0 & \cdots & 0 \end{bmatrix}_{1\times n}$, the estimated acceleration is:
\begin{equation}\label{eq_acc_est}
\hat{\boldsymbol{a}}^b = \textbf{e}_2\hat{\boldsymbol{V}}.
\end{equation}
\hl{In this formulation, we assume that, over the short fitting window, the AUV motion expressed in the body frame can be locally approximated by the polynomial model in}~\eqref{eqn:v_pol}.
\hl{The acceleration covariance} follows directly from \eqref{eq_Vhat}. Since each axis is estimated independently using the same design matrix $\mathbf{\Psi}$, and the noise is isotropic across axes, $\text{Cov}(\hat{\boldsymbol{V}}_{:,i}) = \sigma_v^2 (\mathbf{\Psi}^{\mathsf T}\mathbf{\Psi})^{-1}$ for each axis $i \in \{1,2,3\}$, with zero cross-axis covariance. The scalar per-axis acceleration variance is therefore:
\begin{equation}\label{eq_racc_scalar}
\sigma_{acc}^2 = \sigma_v^2 \, \textbf{e}_2 (\mathbf{\Psi}^{\mathsf T}\mathbf{\Psi})^{-1} \textbf{e}_2^{\mathsf T},
\end{equation}
and the corresponding $3\times3$ measurement noise covariance used as $\mathbf{R}_{acc}$ in the filter update is:
\begin{equation}\label{eq_Racc}
\mathbf{R}_{acc} = \sigma_{acc}^2 \, \mathbf{I}_3.
\end{equation}
\noindent
\hl{Since the newly derived acceleration measurement in}~\eqref{eq_acc_est} \hl{is obtained from a sequence of DVL velocity measurements, including the current velocity measurement, the resulting acceleration and current velocity measurement errors are statistically correlated. This cross-correlation can be obtained directly from the linear acceleration estimator. To this end we define: }
\begin{equation}\label{eqn:cT}
    \boldsymbol{c}^{T}
    =
    \boldsymbol{e}_{2}
    \left(\mathbf{\Psi}^{T}\mathbf{\Psi}\right)^{-1}
    \mathbf{\Psi}^{T}
    =
    \begin{bmatrix}
        c_{0} & c_{1} & \cdots & c_{m-1}
    \end{bmatrix},
\end{equation}
\hl{such that the acceleration estimate in}~\eqref{eq_acc_est} \hl{can be written, using }\eqref{eqn:cT}\hl{, as:}
\begin{equation}
    \hat{\boldsymbol{a}}^b
    =
    \boldsymbol{c}^{T}\mathbf{Y}
    =
    \sum_{i=0}^{m-1} c_i \mathbf{y}_i.
\end{equation}
\hl{Accordingly, the noise component of the acceleration estimate is}
\begin{equation}
    \boldsymbol{\omega}_{a}
    =
    \sum_{i=0}^{m-1} c_i \boldsymbol{\omega}_{v,i}.
\end{equation}
\hl{At each DVL update, the current velocity measurement} $\boldsymbol{}{y}_0$ \hl{is used 
both directly in the conventional velocity update and as part of the 
acceleration estimate. Therefore, the cross-covariance between the current 
DVL velocity measurement noise and the derived acceleration measurement noise is:}
\begin{align}\label{eqn:Eww}
    \mathbf{R}_{va}
    &=
    \mathbb{E}
    \left[
        \boldsymbol{\omega}_{v,0}
        \boldsymbol{\omega}_{a}^{T}
    \right] \nonumber \\
    &=
    \sum_{i=0}^{m-1} c_i
    \mathbb{E}
    \left[
        \boldsymbol{\omega}_{v,0}
        \boldsymbol{\omega}_{v,i}^{T}
    \right].
\end{align}
\hl{Under the temporally white and isotropic DVL measurement-noise assumption 
introduced in}~\eqref{eqn:noise_yk},
\begin{equation}
    \mathbb{E}
    \left[
        \boldsymbol{\omega}_{v,0}
        \boldsymbol{\omega}_{v,i}^{T}
    \right]
    =
    \begin{cases}
        \sigma_v^2 \mathbf{I}_3, & i=0, \\
        \mathbf{0}_{3\times3},   & i\neq0,
    \end{cases}
\end{equation}
\hl{and hence}~\eqref{eqn:Eww} \hl{is rewritten as:}
\begin{equation}
    \mathbf{R}_{va}
    =
    c_0 \sigma_v^2 \mathbf{I}_3,
\end{equation}
\hl{where}
\begin{equation}
    c_0
    =
    \mathbf{e}_{2}
    \left(\mathbf{\Psi}^{T}\mathbf{\Psi}\right)^{-1}
    \mathbf{\Psi}^{T}
    \mathbf{e}_{1}^{T},
\end{equation}
\hl{with} $\mathbf{e}_{1}=[1,0,\ldots,0]$ \hl{selecting the current DVL velocity 
measurement from the fitting window.}
\\ \noindent
\hl{Consequently, when the velocity and acceleration measurements are jointly 
incorporated into the filter, the complete measurement-noise covariance is:}
\begin{equation}
    \mathbf{R}_{\mathrm{DVL}}
    =
    \begin{bmatrix}
        \sigma_v^2 \mathbf{I}_3 & \mathbf{R}_{va} \\
        \mathbf{R}_{va}^{T}     & \mathbf{R}_{\mathrm{acc}}
    \end{bmatrix}
    =
    \begin{bmatrix}
        \sigma_v^2 \mathbf{I}_3
        &
        c_0 \sigma_v^2 \mathbf{I}_3
        \\
        c_0 \sigma_v^2 \mathbf{I}_3
        &
        \sigma_{\mathrm{acc}}^2 \mathbf{I}_3
    \end{bmatrix}.
\end{equation}
\hl{Thus, both the uncertainty propagated from the DVL velocity measurements into 
the acceleration estimate and the statistical dependence between the current 
velocity and acceleration measurements are explicitly accounted for in the 
filter update.}

\subsection{Acceleration-Aided INS/DVL Fusion}
\noindent
In this section, we derive the measurement model for the acceleration update. The measurement residual is defined as:
\begin{equation}\label{eq:acc_res}
\delta\boldsymbol{ z}_a=\boldsymbol{\hat{a}}^{b}-\boldsymbol{\tilde{a}}^{b}
\end{equation}
where $\boldsymbol{\hat{a}}^{b}$ is obtained from \eqref{eq_acc_est} and $\boldsymbol{\tilde{a}}^{b}$ is the corresponding inertial acceleration.\\
The true acceleration, expressed in the body frame, is given by:
\begin{equation} \label{eq: true acc}
\boldsymbol{a}^{b} = \boldsymbol{f}^{b}_{ib} + \mathbf{R}^{b}_{n} \cdot \boldsymbol{g}^{n}
\end{equation}
Since the rotation matrix $\mathbf{R}^{b}_{n}$ is not perfectly known in practice, it is expressed through its estimate and a misalignment correction:
\begin{equation} \label{eq: true dcm}
\mathbf{R}^{b}_{n}=\mathbf{\hat{R}}^{b}_{n}(\mathbf{I}-{\left[\boldsymbol{\phi}^{n}\times \right]})
\end{equation}
Similarly, the true specific force $\boldsymbol{f}^{b}_{ib}$ is related to the measured specific force through the accelerometer bias residual:
\begin{equation} \label{eq: true specific force}
\boldsymbol{f}^b_{ib}=\boldsymbol{\tilde{f}}^{b}_{ib}-\boldsymbol{b}_{a}
\end{equation}
Substituting \eqref{eq: true dcm}-\eqref{eq: true specific force} into the residual definition \eqref{eq:acc_res}-\eqref{eq: true acc} and expanding term by term:
\begin{align}
\boldsymbol{\delta a} &= \boldsymbol{\tilde{a}}^{b} - \boldsymbol{a}^{b} \notag \\[2pt]
&= \left(\boldsymbol{\tilde{f}}^{b}_{ib} + \mathbf{\hat{R}}^{b}_{n}\boldsymbol{g}^{n}\right)
- \left(\boldsymbol{f}^{b}_{ib} + \mathbf{R}^{b}_{n}\boldsymbol{g}^{n}\right) \notag \\[2pt]
&= \left(\boldsymbol{\tilde{f}}^{b}_{ib} + \mathbf{\hat{R}}^{b}_{n}\boldsymbol{g}^{n}\right)
- \left[\left(\boldsymbol{\tilde{f}}^{b}_{ib} - \boldsymbol{b}_{a}\right)
+ \mathbf{\hat{R}}^{b}_{n}\left(\mathbf{I}_{3} - \left[\boldsymbol{\phi}^{n}\times\right]\right)\boldsymbol{g}^{n}\right] \notag \\[2pt]
&= \cancel{\boldsymbol{\tilde{f}}^{b}_{ib}} + \mathbf{\hat{R}}^{b}_{n}\boldsymbol{g}^{n}
- \cancel{\boldsymbol{\tilde{f}}^{b}_{ib}} + \boldsymbol{b}_{a}
- \mathbf{\hat{R}}^{b}_{n}\boldsymbol{g}^{n}
+ \mathbf{\hat{R}}^{b}_{n}\left[\boldsymbol{\phi}^{n}\times\right]\boldsymbol{g}^{n} \notag \\[2pt]
&= \boldsymbol{b}_{a} + \mathbf{\hat{R}}^{b}_{n}\left[\boldsymbol{\phi}^{n}\times\right]\boldsymbol{g}^{n} \notag \\[2pt]
&= \boldsymbol{b}_{a} - \mathbf{\hat{R}}^{b}_{n}\left[\boldsymbol{g}^{n}\times\right]\boldsymbol{\phi}^{n}
\label{eq: dacc}
\end{align}
where the last line follows from the cross-product identity $[\boldsymbol{\phi}\times]\boldsymbol{g} = -[\boldsymbol{g}\times]\boldsymbol{\phi}$. Reading off the coefficients of each error state in \eqref{eq: dacc}, the corresponding measurement matrix $\mathbf{H_a}$ is:
\begin{equation} \label{eq: H_a}
\mathbf{H_a}=\left[ \begin{array}{cccc}
\mathbf{0}_{3\times3} & -\mathbf{\hat{R}}^{b}_{n}\cdot {\left[\boldsymbol{g}^{n} \times \right]} & {\mathbf{I}_{3}} & \mathbf{0}_{3\times3}\end{array} \right].
\end{equation}
To summarize, the acceleration-aided model is defined by the measurement residual \eqref{eq:acc_res} and its associated measurement matrix \eqref{eq: H_a}.

\subsection{Extended DVL Update Model}
\noindent
Regardless of the nonlinear filter employed, the baseline DVL update model relies solely on the velocity measurement, comprising the residual \eqref{eq: z_vel} and measurement matrix \eqref{eq:DVL_vel_H}. Our proposed extended model augments this baseline by incorporating the acceleration measurement, adding the residual \eqref{eq:acc_res} and measurement matrix \eqref{eq: H_a}. Stacking \eqref{eq: z_vel} and \eqref{eq:acc_res} together, the extended DVL measurement residual is:
\begin{equation} \label{eq:DVL_E_Res}
\delta\boldsymbol{z}_{DVL}= \left[ \begin{array}{c}
\boldsymbol{\hat{v}}^{b} - \boldsymbol{\tilde{v}}_{DVL}^{b} \\
\boldsymbol{\hat{a}}^{b}-\boldsymbol{\tilde{a}}^{b}
\end{array} \right ] 
\end{equation}
\noindent
and the corresponding measurement matrix follows by stacking \eqref{eq:DVL_vel_H} and \eqref{eq: H_a}:
 \begin{equation} \label{eq:DVL_E_H}
 \mathbf{H}_{DVL} = \left[ \begin{array}{cccc}
\mathbf{\hat{R}}_{n}^{b} & -\mathbf{\hat{R}}_{n}^{b}
[\boldsymbol{\tilde{v}}_{DVL}^{n} \times] & \mathbf{0}_{3\times3} & \mathbf{0}_{3\times3} \\
\mathbf{0}_{3\times3} & -\mathbf{\hat{R}}^{b}_{n}\cdot {\left[\boldsymbol{g}^{n} \times \right]} & {\mathbf{I}_{3}} & \mathbf{0}_{3\times3}
\end{array} \right]. 
\end{equation}
\noindent
Together, \eqref{eq:DVL_E_Res} and \eqref{eq:DVL_E_H} define the extended DVL update applied to the navigation filter.
\subsection{Observability Analysis} \label{subsec: obsv}
\noindent To assess the observability contribution of the DVL-based acceleration measurement, we analytically derive its unobservable subspace using an observability Gramian approach \cite{maybeck1982stochastic}, following a procedure similar to \cite{klein2015observability}. The unobservable subspace is obtained from the right null space of the Gramian matrix, identifying the state directions that cannot be directly observed or estimated. This is expressed as:
\begin{equation} \label{eq: Gramian}
\boldsymbol{\mathrm{H_a}}(t)\mathbf{\Phi}(t, t_{0})u_{0}=\boldsymbol{0}_{M}
\end{equation}
where $M$ is the number of measurements per time step, $u_{0}$ denotes a set of solutions spanning the unobservable subspace of the state vector, $\mathbf{H_a}$ is the measurement matrix defined in \eqref{eq: H_a}, and $\mathbf{\Phi}(t, t_{0})$ is the state transition matrix \cite{klein2015observability}:
\begin{equation} \label{eq: Phi closed form}
\mathbf{\Phi}(t, t_{0}) =\left[ \begin{array}{cccc}
\mathbf{I_{3}} & \mathbf{S_{t}} & \mathbf{R_{t}} & \mathbf{M_{t}} \\ 
\mathbf{0_{3\times3}} &\mathbf{I_{3}} & \mathbf{0_{3\times3}} & \mathbf{R_{t}} \\ 
\mathbf{0_{3\times3}} & \mathbf{0_{3\times3}} & \mathbf{I_{3}} & \mathbf{0_{3\times3}} \\ 
\mathbf{0_{3\times3}} & \mathbf{0_{3\times3}} & \mathbf{0_{3\times3}} &\mathbf{I_{3}} \end{array}
\right]
\end{equation}
with
\begin{equation}
\begin{aligned}
\mathbf{M_{t}} = -\int_{t_{0}}^{t}[\boldsymbol{f}^{n}(s)\times]\mathbf{R_{s}}\,ds\\
\mathbf{R_{t}} = \int_{t_{0}}^{t}\mathbf{R}^{n}_{b}(\tau)\,d\tau\\
\mathbf{S_{t}} = \int_{t_{0}}^{t}[\boldsymbol{f}^{n}(\tau)\times]\mathbf{R_{s}}\,d\tau
\end{aligned} .
\end{equation}
The unobservable subspace of the state vector is written as:
\begin{equation} \label{eq: u_0}
\boldsymbol{u}_{0}=\left[ \begin{array}{cccc}
\boldsymbol{u}_{1}^{T} & \boldsymbol{u}_{2}^{T} & \boldsymbol{u}_{3}^{T} & \boldsymbol{u}_{4}^{T}\end{array} \right].
\end{equation}
For tractability, we assume throughout the remainder of the analysis that the navigation and body frames coincide, i.e., $\mathbf{R}^{n}_{b}=\mathbf{I}_{3}$. Since \eqref{eq: Gramian} holds for any time interval, we evaluate it at $t = 0$ and substitute \eqref{eq: H_a}, \eqref{eq: Phi closed form}, and \eqref{eq: u_0} into \eqref{eq: Gramian}:
\begin{equation} \label{eq:obs1}
0 \cdot \boldsymbol{u}_{1}^{T}
-[\boldsymbol{g}\times]\cdot \boldsymbol{u}_{2}^{T}
+\mathbf{I}_{3}\cdot \boldsymbol{u}_{3}^{T}
-[\boldsymbol{g}\times]\cdot \mathbf{R}_{t}\cdot \boldsymbol{u}_{4}^{T}
=0.
\end{equation}
\noindent
\hl{Evaluating}~\eqref{eq:obs1} \hl{at $t=t_0$, for which
$\mathbf{R}_{t}=\mathbf{0}$, yields}
\begin{equation} \label{eq:obs2}
\mathbf{I}_{3}\cdot \boldsymbol{u}_{3}^{T}
=
[\boldsymbol{g}\times]\cdot \boldsymbol{u}_{2}^{T}.
\end{equation}
\noindent
\hl{Since}~\eqref{eq:obs1} \hl{must hold for all $t$, differentiating it with
respect to time provides the additional condition}
\begin{equation} \label{eq:obs3}
-[\boldsymbol{g}\times]\cdot \boldsymbol{u}_{4}^{T}=0.
\end{equation}

\noindent 
Combining \eqref{eq:obs2} and \eqref{eq:obs3} gives the unobservable subspace matrix:
\begin{equation}
\mathbf{U} =
\begin{bmatrix}
\boldsymbol{0}_{1\times3} &
\boldsymbol{0}_{1\times3} &
\boldsymbol{0}_{1\times3} &
0 & 0 & 1 \\
\mathbf{0}_{3\times3} &
\mathbf{I}_{3} &
-[\boldsymbol{g}\times] &
\mathbf{0}_{3\times3}
\end{bmatrix}.
\end{equation}
This subspace is identical to the unobservable subspace of the DVL velocity measurement reported in \cite{klein2015observability}. The DVL acceleration measurement therefore does not extend the observable state space beyond what velocity measurements already provide. Nonetheless, as shown in the next section, incorporating this additional measurement is still expected to improve the overall accuracy of the estimated states.
\section{Analysis and Results}\label{sec:results}
\subsection{Dataset Acquisition}\label{sec:DA}
\begin{figure}[h!]
    \centering
    \includegraphics[width=\columnwidth]{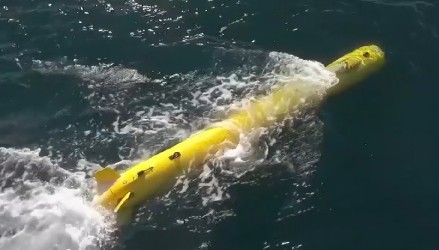}
    \caption{The Snapir AUV during a field deployment at sea.}
    \label{fig:auv_deployment}
\end{figure}

\begin{figure}[h!]
    \centering
    \includegraphics[width=\columnwidth]{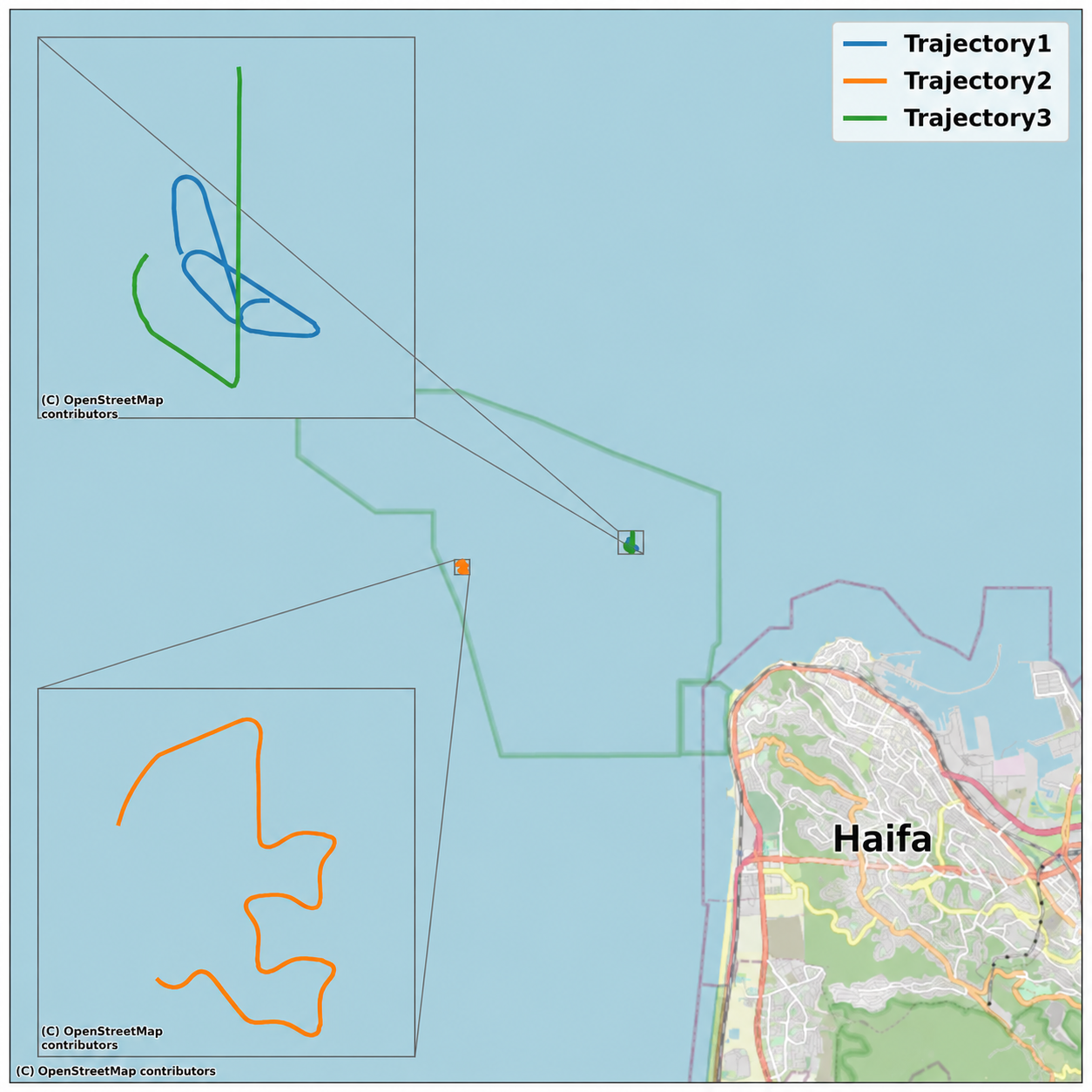}
    \caption{The three recorded trajectories overlaid on the survey area near Rosh Carmel, Haifa. Insets show zoomed-in views of Trajectory 1 and 3 (top) and Trajectory 2 (bottom).}
    \label{fig:trajectories_map}
\end{figure}

\noindent
Data were collected at sea near the Rosh Carmel nature reserve, off the Haifa coast, Israel, using the Snapir AUV, a deep-water variant of the ECA Group A18D platform rated for missions down to 3000 m with an endurance of roughly 21 hours \cite{haifa2026oceaninstruments,ECA_AUV}. Navigation instrumentation onboard consists of an iXblue Phins Subsea fiber-optic gyroscope (FOG) INS, sampling at 100 Hz~\cite{iXblue_PHINS}, paired with a Teledyne RDI Work Horse Navigator DVL operating at 1 Hz with a velocity measurement precision of 0.02 m/s~\cite{TeledyneMarine_DVL}. Fig.~\ref{fig:auv_deployment} shows the vehicle during deployment.
\noindent Three 400 seconds trajectories were selected from the recorded missions, each reflecting a different maneuvering style commonly encountered during AUV operations. The first traces a broad, sweeping loop with gentle curvature, resembling repeated station-keeping or survey-pattern maneuvers. The second follows a serpentine, back-and-forth weaving path, typical of lawnmower-style seabed scanning missions. The third consists of a long straight leg followed by a turn, characteristic of long-range transit or perimeter-following missions. Together, these three profiles capture a representative cross-section of the dynamics an AUV encounters in the field, from tight maneuvering to sustained, gentle turns, providing a varied basis for evaluating the proposed approach. The recorded trajectories, projected onto the survey area, are shown in Fig.~\ref{fig:trajectories_map}. \hl{The data are publicly available in the} \href{https://github.com/ansfl/A-KIT/tree/main/Data}{\hl{A-KIT dataset repository}}\hl{, and an in-depth description of the dataset can be found in}~\cite{cohen2025adaptive}.

\subsection{Data Augmentation}
\noindent
Since the Snapir AUV is equipped with a navigation-grade FOG-based INS, its raw measurements are considerably cleaner than those typically produced by the lower-cost IMUs found on smaller or budget AUV platforms. To assess the applicability of the proposed approach under such conditions, we synthetically degraded (yet preserve the real-world behavior) the IMU measurements while leaving the DVL data untouched, under the assumption that both sensors are otherwise well calibrated. Independent zero-mean Gaussian noise was added to each axis of the gyroscope and accelerometer signals, with a standard deviation consistent with a low-cost, consumer-grade MEMS IMU rather than the vehicle's native navigation-grade sensor. The injected noise parameters are summarized in Table~\ref{tab:sensor_degradation}. This procedure allows the IMU measurements to emulate those of a lower-cost MEMS sensor while preserving the same underlying trajectories and DVL observations.

\begin{table}[h!]
\centering
\caption{Per-axis white noise standard deviation injected into the IMU measurements to emulate a low-cost MEMS-grade sensor.}
\label{tab:sensor_degradation}
\begin{tabular}{lc}
\hline
\textbf{Signal} & \textbf{Injected noise std ($\sigma$)} \\
\hline
Gyroscope & $10^{-2}$ rad/s \\
Accelerometer & $10^{-1}$ m/s$^2$ \\
\hline
\end{tabular}
\end{table}
\subsection{Evaluation Metrics}
\noindent
To quantify the performance of the proposed acceleration-aided filter, we adopt the root mean square error (RMSE) as the primary accuracy metric. For a generic quantity of interest $\boldsymbol{\xi}$, with true value $\boldsymbol{\xi}_k$ and estimated value $\boldsymbol{\hat{\xi}}_k$ at time step $k$, the RMSE is defined as:
\begin{equation}
    \text{RMSE}(\boldsymbol{\xi}) = \sqrt{\frac{1}{T}\sum_{k=1}^{T}
    \left\|\boldsymbol{\xi}_k - \boldsymbol{\hat{\xi}}_k\right\|^2}
    \label{eqn:rmse}
\end{equation}
where $T$ is the total number of time steps. In the following, this metric is applied to $\boldsymbol{\xi} \in \{\boldsymbol{p}, \boldsymbol{v}, \boldsymbol{\phi}\}$, corresponding to position [m], velocity [m/s], and orientation [rad], respectively, allowing the contribution of the acceleration update to be assessed across each component of the navigation state.
\\ \noindent
Beyond point-estimate accuracy, we are also interested in whether incorporating the DVL-derived acceleration measurement reduces the estimator's uncertainty relative to the baseline. To this end, we define the percent covariance improvement (PCI) as the relative reduction in the trace of the error-state covariance of the proposed extended filter with respect to the baseline EKF, which uses only velocity updates~\cite{simon2006optimal}:
\begin{equation}
    \text{PCI}_{j} = 100\times\frac{\text{Tr}\!\left(
    \mathbf{P}_{j}^{\text{base}} - \mathbf{P}_{j}^{\text{ext}} \right)}
    {\text{Tr}\!\left(\mathbf{P}_{j}^{\text{base}}\right)}
    \label{eqn:pci}
\end{equation}
where $\mathbf{P}_{j}^{\text{base}}$ is the \textit{a-posteriori} error-state covariance of the baseline EKF, using only the DVL velocity update \eqref{eq: z_vel}-\eqref{eq:DVL_vel_H}, and $\mathbf{P}_{j}^{\text{ext}}$ is the \textit{a-posteriori} error-state covariance of the proposed extended EKF, incorporating the combined velocity-acceleration update \eqref{eq:DVL_E_Res}-\eqref{eq:DVL_E_H}. A positive PCI value indicates that the acceleration update reduces the estimator's uncertainty relative to the velocity-only baseline, while a value near zero or negative would suggest no meaningful benefit from the added measurement.
\subsection{Results and Discussion}
\begin{figure}[h!]
    \centering
     \includegraphics[width=\columnwidth]{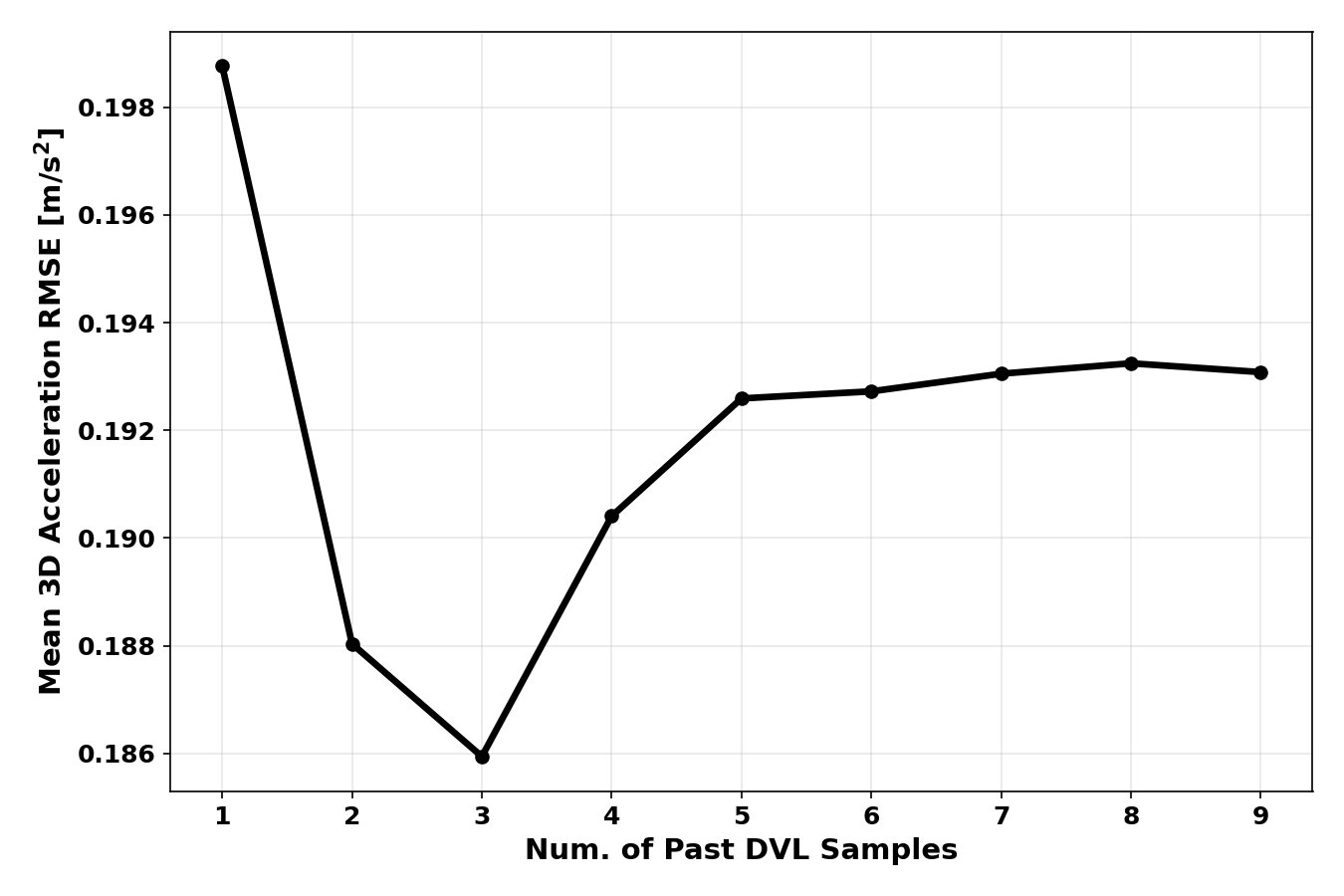}
    \caption{Acceleration RMSE as a function of the number of past DVL velocity measurements used in the polynomial fit.} 
    \label{fig:acc_rmse_window}
\end{figure}

\noindent
The proposed acceleration-aided EKF (AA-EKF) is evaluated against the baseline EKF, which relies solely on DVL velocity updates, across three test trajectories exhibiting distinct maneuvering profiles. We first examine the sensitivity of the acceleration estimate to the number of past DVL velocity samples used in the polynomial fit, followed by a trajectory-level, aggregate, and covariance-level analysis of the resulting navigation performance.
\\ \noindent 
The original dataset presented in~\cite{cohen2025adaptive} includes 13 trajectories. while in this work we use three of them for testing, as addressed in Section~\ref{sec:DA}. However, to determine the ideal number of past DVL measurements, the remaining trajectories were used instead, so that no bias would be introduced into the test results. Fig.~\ref{fig:acc_rmse_window} shows the acceleration RMSE as a function of the number of past DVL velocity measurements used in the polynomial fitting window described in Section~\ref{sec:proposed}.

\noindent

\begin{figure*}[h!]
    \centering
    \begin{subfigure}[b]{0.32\textwidth}
        \centering
        \includegraphics[width=\textwidth]{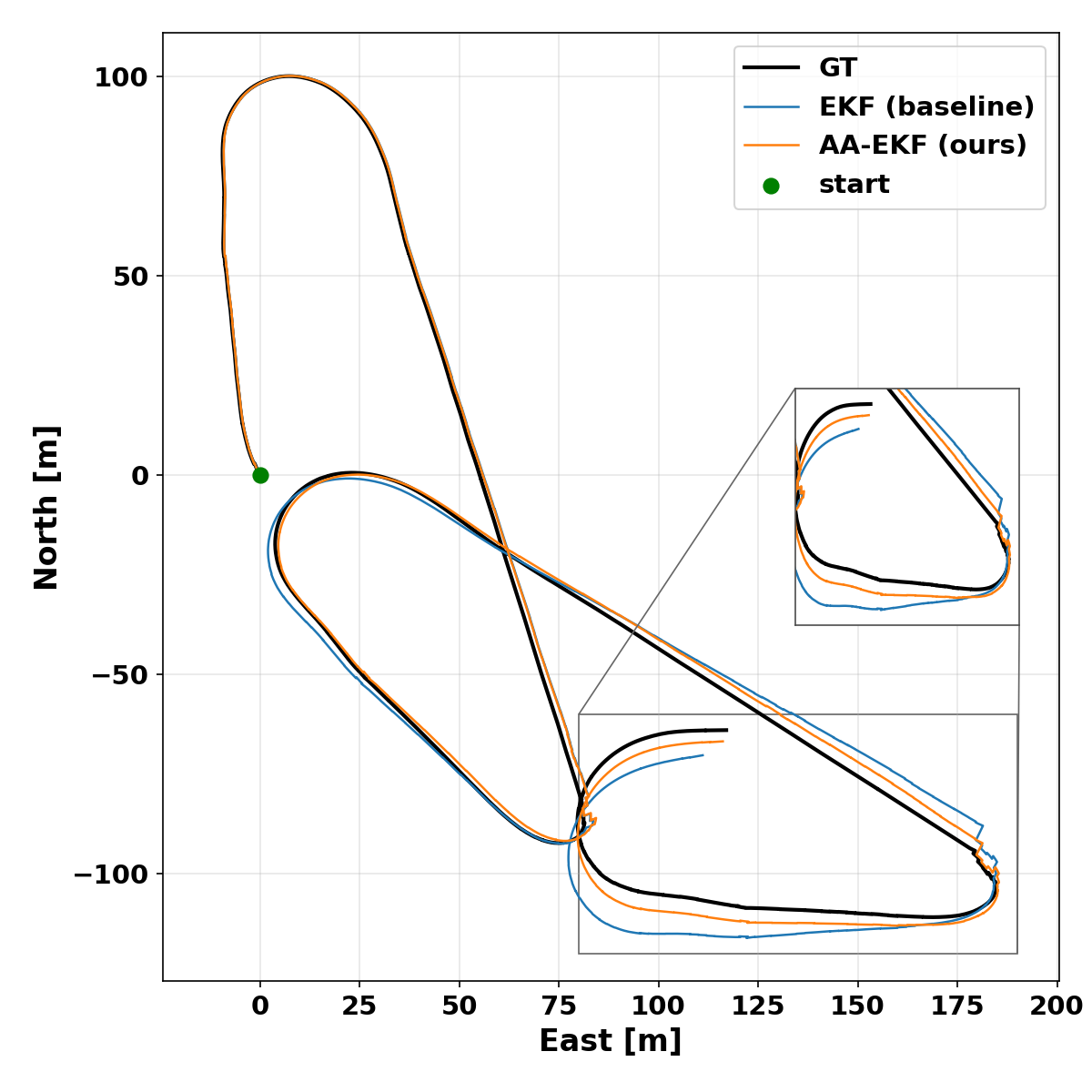}
        \caption{Trajectory 1}
        \label{fig:traj1_2d}
    \end{subfigure}
    \hfill
    \begin{subfigure}[b]{0.32\textwidth}
        \centering
        \includegraphics[width=\textwidth]{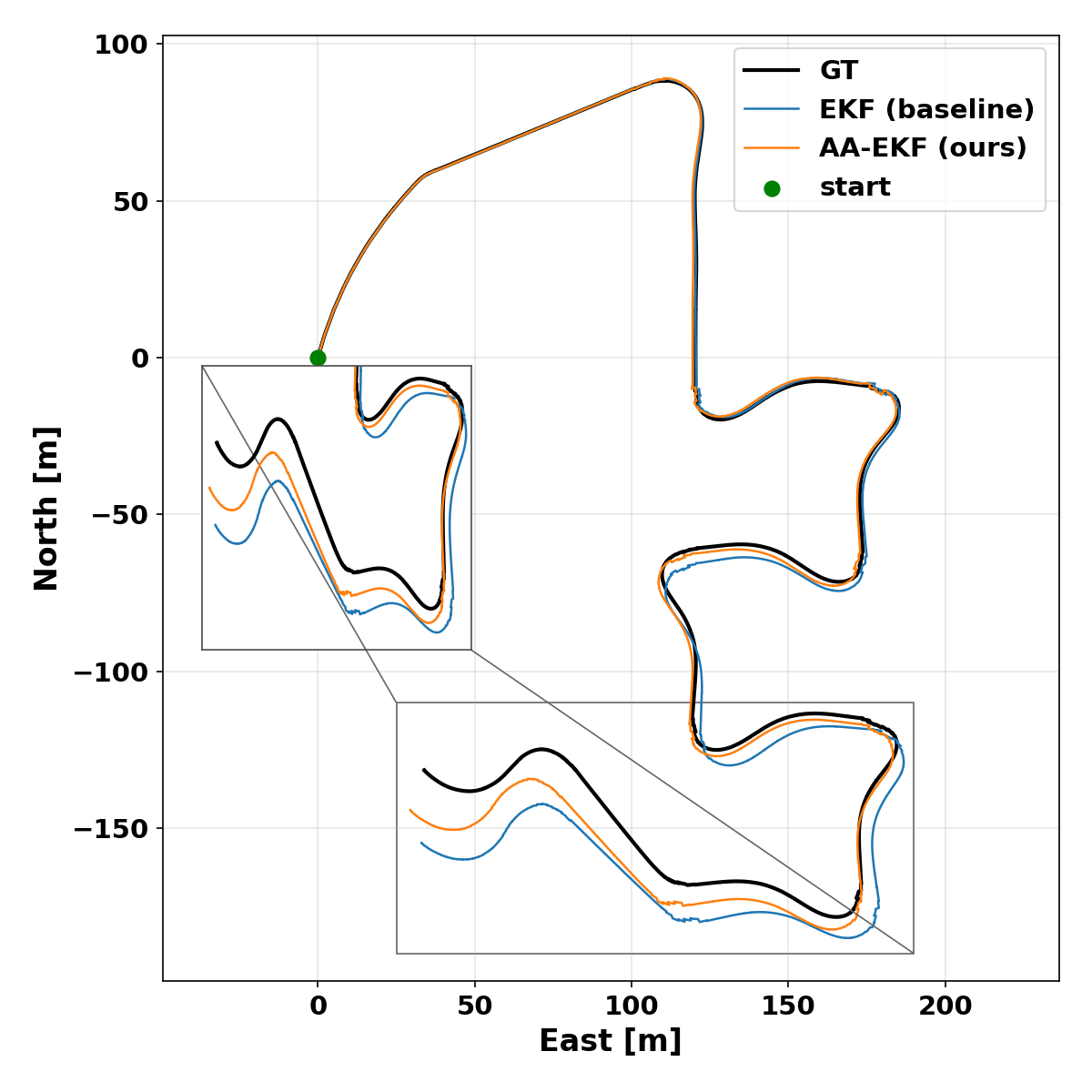}
        \caption{Trajectory 2}
        \label{fig:traj2_2d}
    \end{subfigure}
    \hfill
    \begin{subfigure}[b]{0.32\textwidth}
        \centering
        \includegraphics[width=\textwidth]{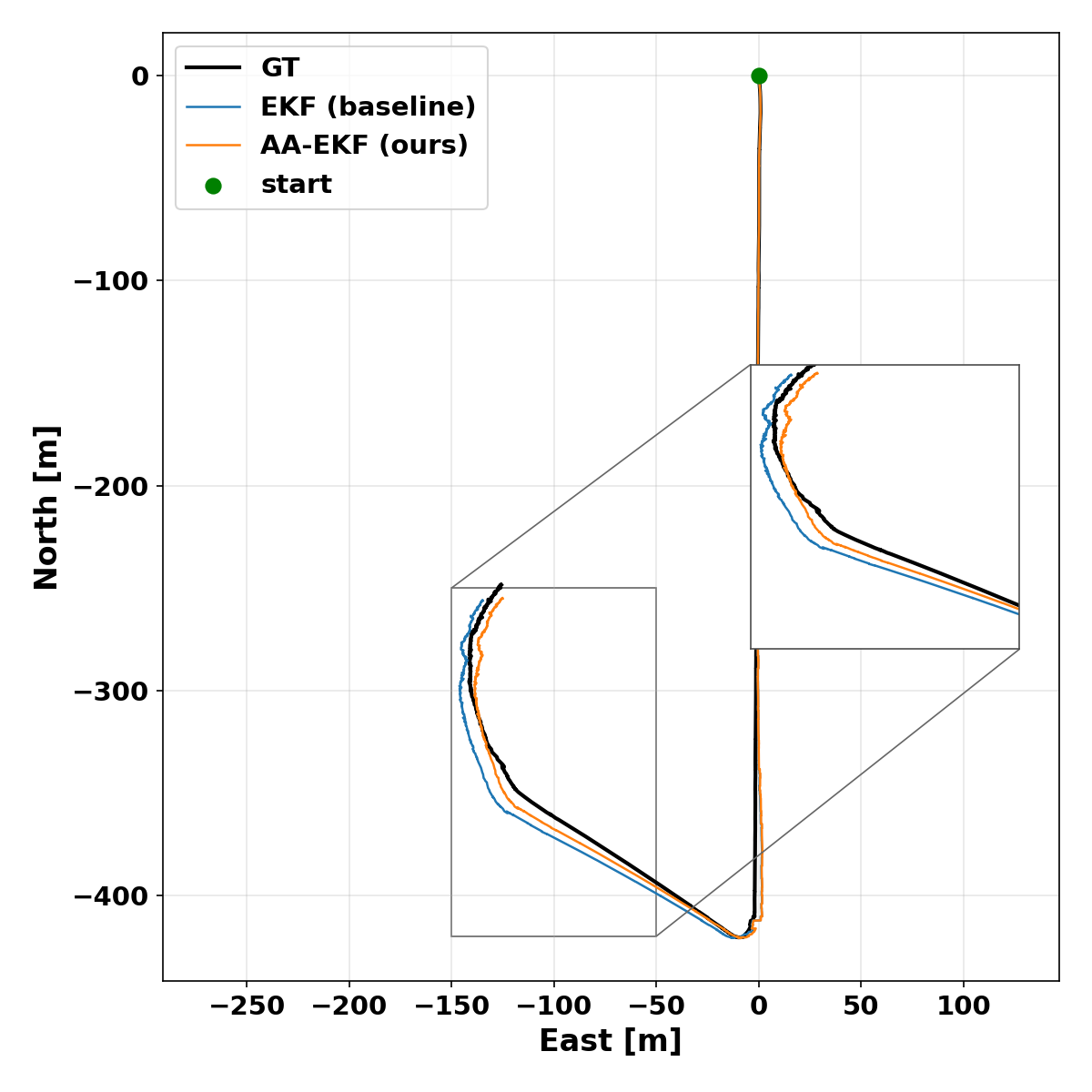}
        \caption{Trajectory 3}
        \label{fig:traj3_2d}
    \end{subfigure}
    \caption{\hl{2D trajectory comparison between the GT, baseline EKF, and AA-EKF (ours) for the three test trajectories.}}
    \label{fig:traj_2d_all}
\end{figure*}

\noindent 
\hl{The outcome suggests that three past DVL measurements, in addition to
the current measurement, provide the lowest acceleration RMSE. Accordingly,
these parameters were used throughout the evaluated trajectories and are
summarized in Table}~\ref{tab:poly_params}.
\begin{table}[h!]
\centering
\caption{\hl{Polynomial fitting parameters used in the experiments.}}
\label{tab:poly_params}
\begin{tabular}{lc}
\hline
Parameter & Value \\
\hline
Retained polynomial terms, $n$ & 2 \\
Past DVL measurements, $m$ & 3 \\
Total DVL measurements & 4 \\
DVL sampling frequency & 1 [Hz] \\
Measurement timestamps & $t_k,\;t_k-1,\;t_k-2,\;t_k-3$ [s] \\
Fitting-window duration & 3 [s] \\
\hline
\end{tabular}
\end{table}
Table~\ref{tab:rmse_results} summarizes the position, velocity, and orientation RMSE obtained by the baseline EKF and the proposed AA-EKF across the three evaluated trajectories, along with the relative improvement achieved by incorporating the acceleration update.

\begin{figure}[t!]
    \centering
    \includegraphics[width=\columnwidth]{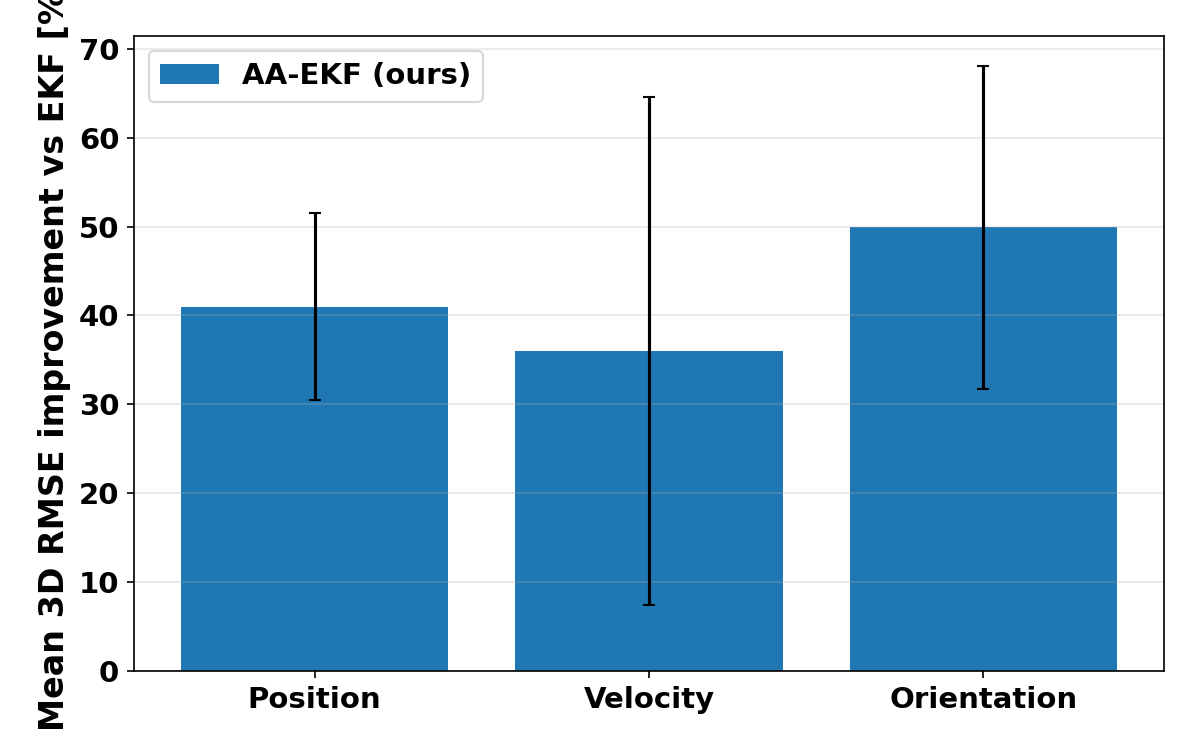}
    \caption{\hl{Mean relative improvement in 3D RMSE of the AA-EKF over the baseline EKF across the three test trajectories, for orientation, position, and velocity. Error bars denote the standard deviation across trajectories.}}
    \label{fig:agg_improve}
\end{figure}

\begin{table}[h!]
\centering
\caption{\hl{Position, velocity, and orientation RMSE for the baseline EKF and the proposed AA-EKF across the three test trajectories. Performance improvements relative to the baseline are shown in parentheses.}}
\label{tab:rmse_results}
\resizebox{\columnwidth}{!}{%
\begin{tabular}{ccccc}
\hline
\textbf{Traj.} & \textbf{Method} & \textbf{Pos. [m]} & \textbf{Vel. [m/s]} & \textbf{Orient. [deg]} \\
\hline
\multirow{2}{*}{1} & EKF (baseline) & 4.43 & 0.143 & 3.78 \\
                   & AA-EKF (ours)  & 2.16 (51.3\%) & 0.055 (61.5\%) & 1.24 (67.3\%) \\
\hline
\multirow{2}{*}{2} & EKF (baseline) & 7.96 & 0.136 & 3.69 \\
                   & AA-EKF (ours)  & 4.66 (41.4\%) & 0.080 (41.2\%) & 1.80 (51.3\%) \\
\hline
\multirow{2}{*}{3} & EKF (baseline) & 7.29 & 0.176 & 4.41 \\
                   & AA-EKF (ours)  & 5.08 (30.2\%) & 0.166 (5.2\%) & 3.04 (31.1\%) \\
\hline
\end{tabular}%
}
\end{table}
\noindent
Across all three trajectories, the AA-EKF consistently outperforms the baseline EKF in position, velocity, and orientation accuracy, with 3D position RMSE improvements ranging from 27.5\% to 50.5\% and orientation RMSE improvements between 31.4\% and 64.9\%. The most pronounced gains are observed for Trajectory 1, where the acceleration update reduces position error by roughly half and orientation error by nearly two-thirds. Trajectory 3, in contrast, shows the smallest velocity improvement (5.4\%), suggesting that the benefit of the acceleration update is not uniform across maneuvering styles and may depend on the excitation richness of the trajectory's dynamics.
\begin{figure}[t!]
    \centering
    \includegraphics[width=\columnwidth]{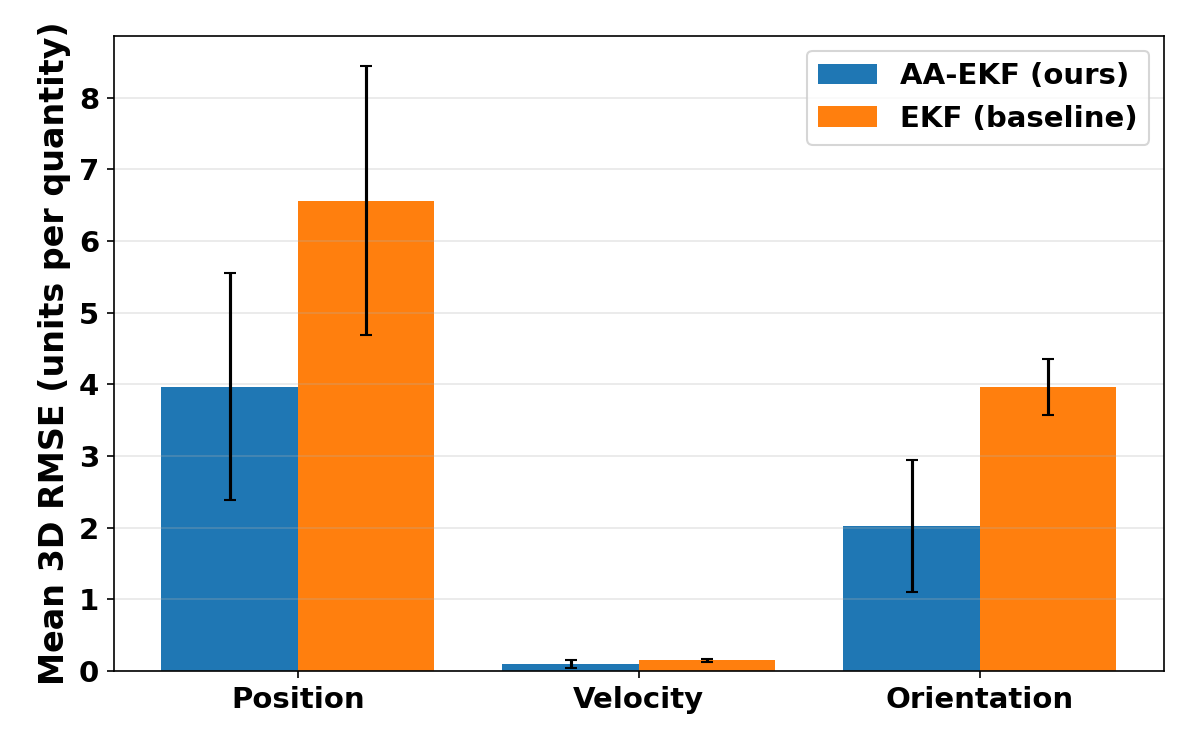}
    \caption{\hl{Mean 3D RMSE across the three test trajectories for the baseline EKF and the proposed AA-EKF, for orientation, position, and velocity. Error bars denote the standard deviation across trajectories.}}
    \label{fig:agg_abs}
\end{figure}
The corresponding 2D trajectories are shown in Fig.~\ref{fig:traj_2d_all}. In all three cases, the AA-EKF visibly tracks the ground truth (GT) more closely than the baseline EKF, particularly during the sharper turning segments highlighted in the zoomed-in insets. For Trajectory 3 (Fig.~\ref{fig:traj3_2d}), the two filters are nearly indistinguishable along the long straight leg, but the AA-EKF noticeably reduces the lateral drift accumulated during the tight turning maneuver. A similar pattern is evident for Trajectory 2 (Fig.~\ref{fig:traj2_2d}), where the baseline EKF drifts progressively further from the GT along the weaving path, while the AA-EKF remains closer to the reference throughout. Trajectory 1 (Fig.~\ref{fig:traj1_2d}) exhibits the most challenging dynamics among the three, with sharp direction reversals. Here, both filters deviate more visibly from the GT than in the other cases.
\\ \noindent
To summarize performance across all three trajectories, Figs.~\ref{fig:agg_abs} and~\ref{fig:agg_improve} present the mean 3D RMSE and the mean relative improvement, respectively, for position, velocity, and orientation, with error bars denoting the standard deviation across trajectories. As shown in Fig.~\ref{fig:agg_abs}, the AA-EKF achieves a lower mean RMSE than the baseline EKF for all three quantities, with the largest absolute reduction observed for position. Fig.~\ref{fig:agg_improve} expresses these same gains as a relative improvement, showing a mean improvement of approximately 50\% for orientation, 40\% for position, and 36\% for velocity.

\begin{figure*}[h!]
    \centering
    \begin{subfigure}[b]{0.32\textwidth}
        \centering
        \includegraphics[width=\textwidth]{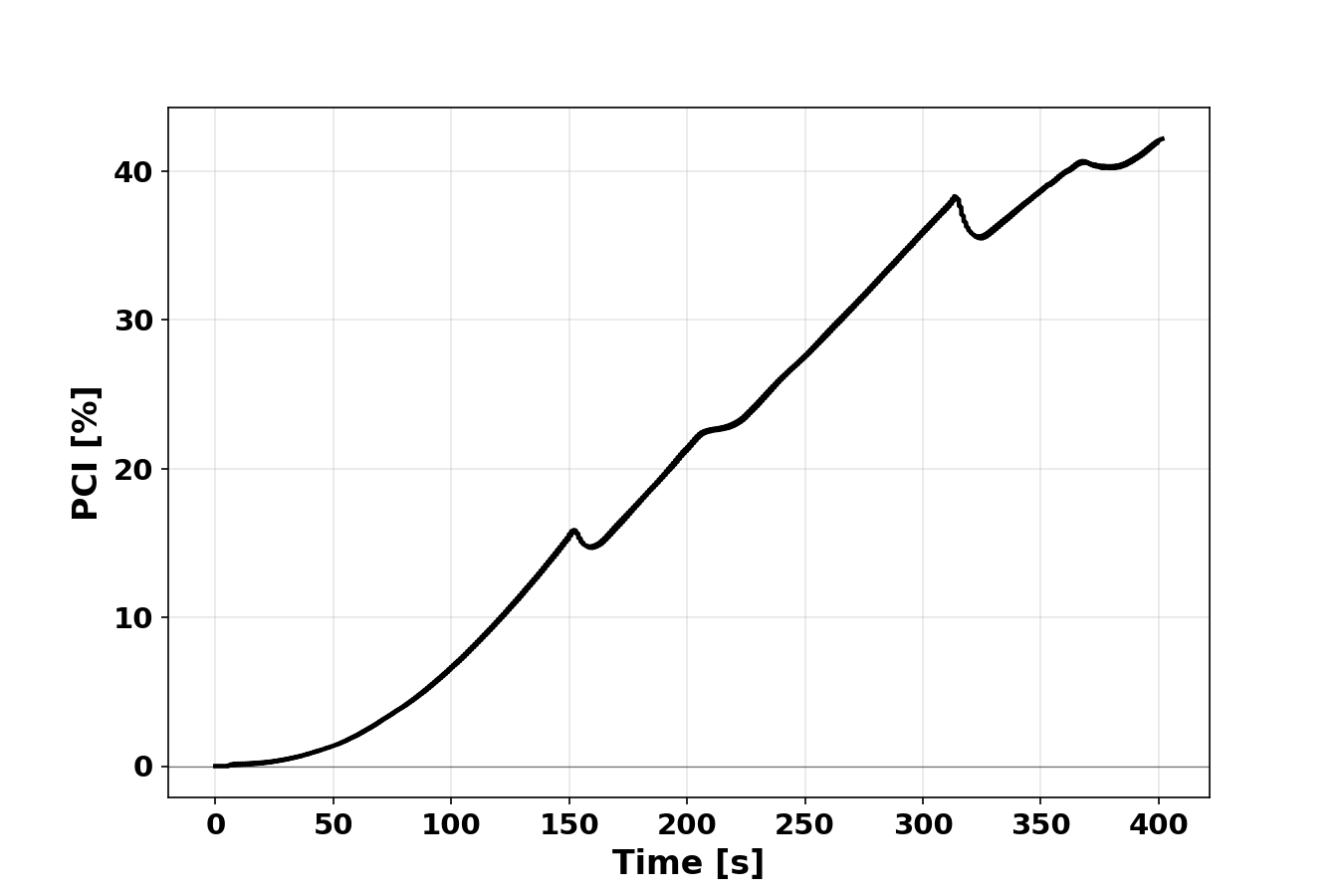}
        \caption{Trajectory 1}
        \label{fig:pci1}
    \end{subfigure}
    \hfill
    \begin{subfigure}[b]{0.32\textwidth}
        \centering
        \includegraphics[width=\textwidth]{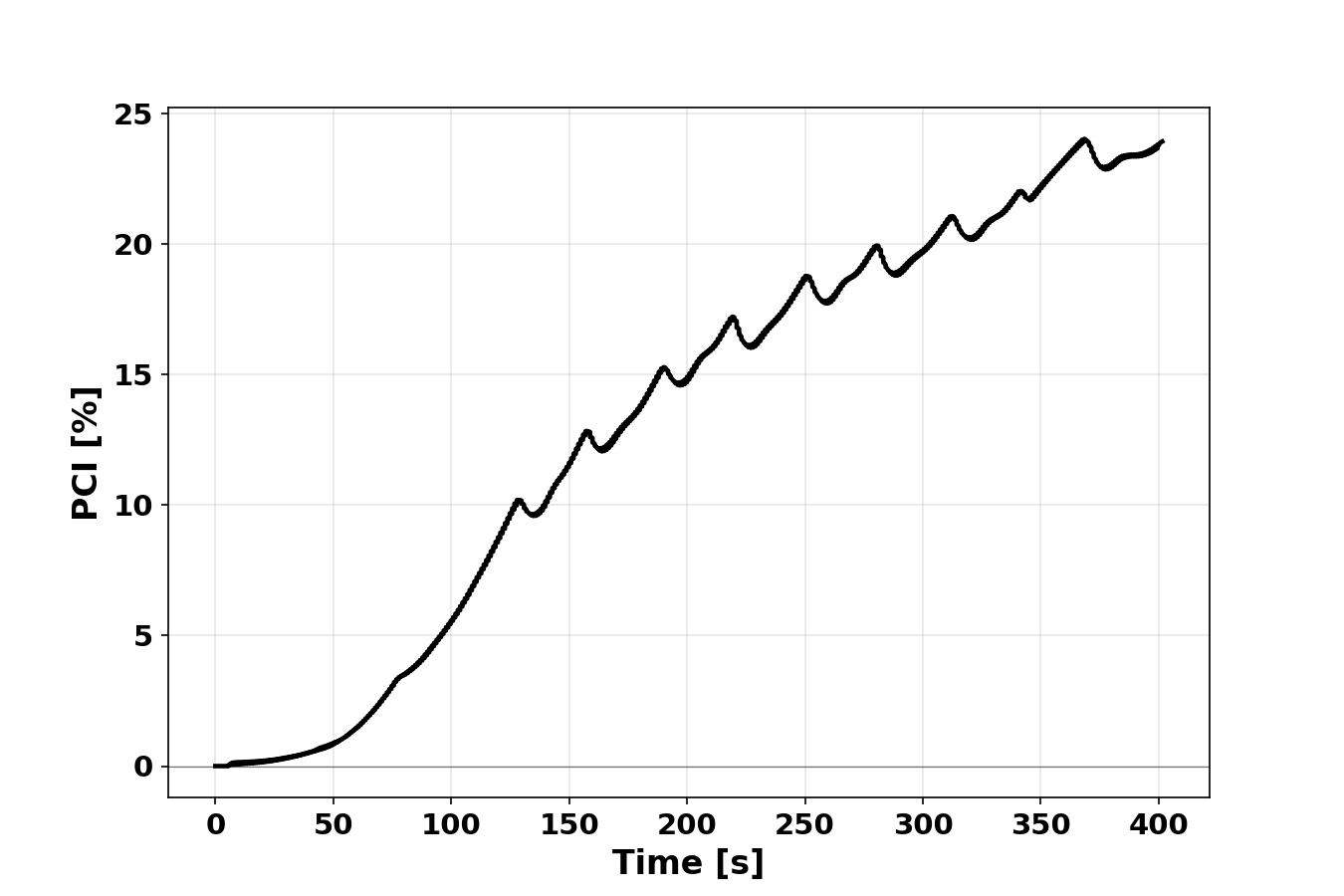}
        \caption{Trajectory 2}
        \label{fig:pci2}
    \end{subfigure}
    \hfill
    \begin{subfigure}[b]{0.32\textwidth}
        \centering
        \includegraphics[width=\textwidth]{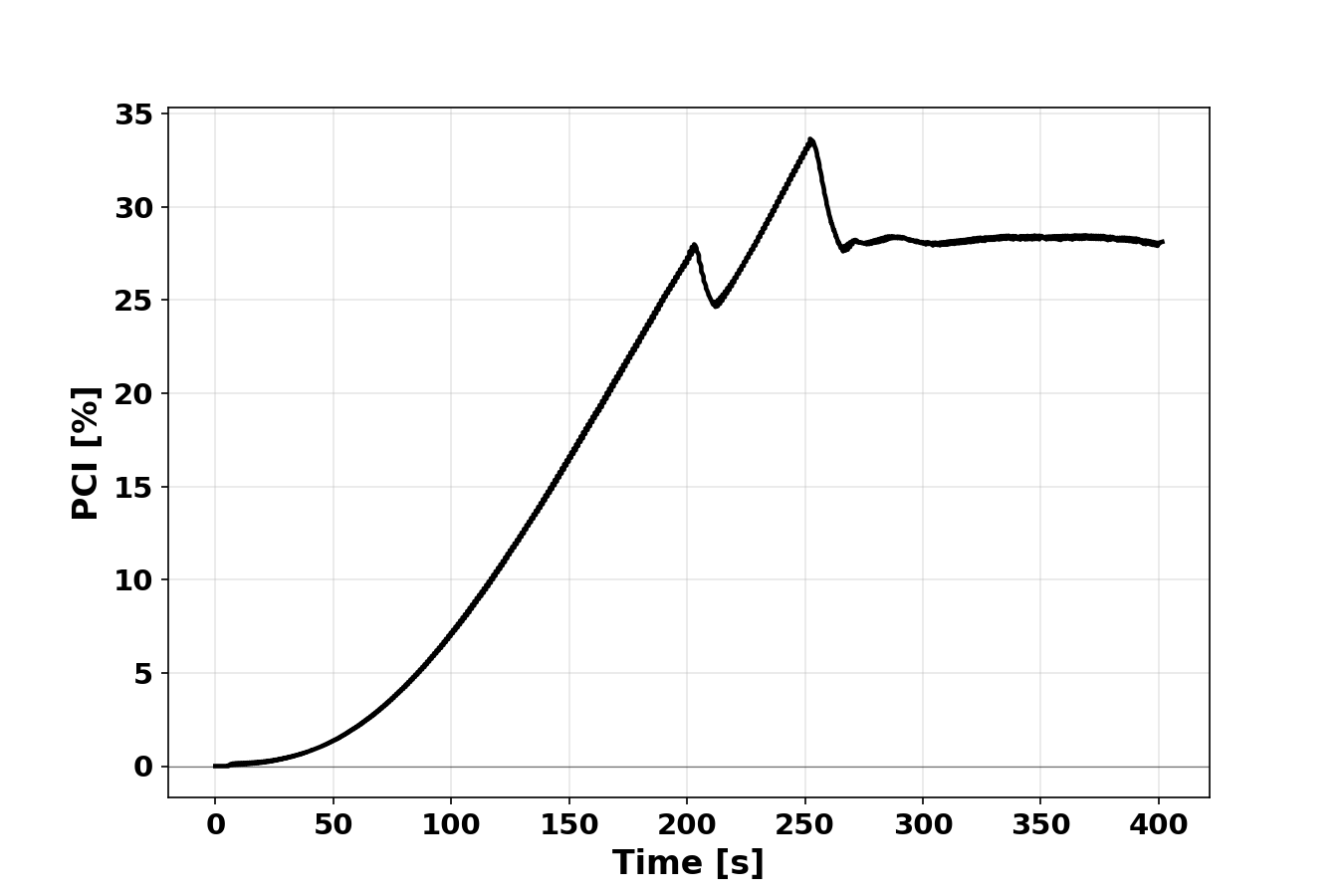}
        \caption{Trajectory 3}
        \label{fig:pci3}
    \end{subfigure}
    \caption{\hl{PCI of the AA-EKF relative to the baseline EKF for the three test trajectories.}}
    \label{fig:pci_all}
\end{figure*}
\noindent
To complement the point-estimate accuracy results, Fig.~\ref{fig:pci_all} presents the PCI, as defined in \eqref{eqn:pci}, for each of the three trajectories. In all cases, the PCI is positive and grows over the course of the trajectory, confirming that the added acceleration update consistently reduces the estimator's uncertainty relative to the velocity-only baseline, rather than merely shifting the point estimate.
For Trajectory 1 (Fig.~\ref{fig:pci1}), the PCI rises steadily to over 40\% by the end of the run, with brief plateaus coinciding with the turning segments identified in Fig.~\ref{fig:traj1_2d}. Trajectory 2 (Fig.~\ref{fig:pci2}) shows a similar monotonic trend, reaching approximately 24\%, consistent with its comparatively smaller RMSE improvement relative to Trajectory 1. Trajectory 3 (Fig.~\ref{fig:pci3}) exhibits a markedly different profile: the PCI increases sharply during the first 250 s, peaking near 33\%, before settling to a steady value around 28\% for the remainder of the run.

\noindent
Taken together, the RMSE, aggregate, and PCI results indicate that incorporating the DVL-derived acceleration measurement consistently improves both the accuracy and the estimated confidence of the navigation solution across trajectories with markedly different dynamics, while the magnitude of this improvement depends on the specific maneuvering profile encountered.
\section{Conclusion}\label{sec:conclude}
\noindent
This paper presented an acceleration-aided extension to the conventional INS/DVL error-state EKF for AUV navigation in GNSS-denied underwater environments. Rather than relying solely on DVL velocity updates, the proposed approach derives an acceleration estimate from a polynomial fit to a short window of past DVL velocity measurements, and incorporates it as an additional measurement update within the same filter framework. An observability analysis showed that this acceleration update does not enlarge the unobservable subspace of the system beyond that of the baseline velocity-only filter, so any accuracy gains stem from the added measurement information rather than from resolving previously unobservable states.
\\ \noindent
The proposed approach was evaluated on three sea-trial trajectories spanning distinct maneuvering profiles, from tight repeated turns to broad sweeping paths, using the Snapir AUV operating in the Mediterranean Sea. Across all three trajectories, the acceleration-aided EKF consistently reduced position, velocity, and orientation RMSE relative to the baseline EKF, with mean improvements of approximately 40\%, 36\%, and 50\%, respectively. The percent covariance improvement metric further confirmed that these accuracy gains were accompanied by a genuine reduction in estimator uncertainty rather than a shift in the point estimate alone.
\\ \noindent
Despite these consistent improvements, several limitations should be acknowledged. First, the magnitude of the improvement varied noticeably across trajectories which suggests the benefit of the acceleration update is sensitive to the excitation richness of the vehicle's motion, and may be more limited during rapid, high-curvature maneuvers. Second, the polynomial acceleration model assumes local smoothness in the velocity signal over the fitting window, an assumption that degrades if the true acceleration changes abruptly within that window, so the choice of window length constitutes a trade-off between noise suppression and responsiveness.
\\ \noindent
Beyond the reported accuracy gains, the strength of this approach lies in its accessibility: since it relies only on information already produced by sensors already onboard, it can be retrofitted onto existing AUVs through a firmware update alone, without new procurement, recalibration, or hardware qualification cycles. This lowers the barrier to deploying more reliable navigation across AUVs, including lower-cost platforms that most need such improvements, supporting the broader push toward affordable, scalable underwater exploration and monitoring.
\section*{Acknowledgment}
\noindent
N.C. is supported by the Maurice Hatter Foundation and the University of Haifa presidential scholarship for outstanding students on a direct Ph.D. track.

\bibliographystyle{IEEEtran}
\bibliography{bio.bib}

\end{document}